\documentclass[english,aps,prx,twocolumn,superscriptaddress,longbibliography]{revtex4-2}
\usepackage{amssymb}
\usepackage{amsmath}
\usepackage{graphicx}
\usepackage{natbib}
\usepackage{epstopdf}
\usepackage{color}
\usepackage{braket}
\usepackage{physics}
\usepackage{mathrsfs}
\usepackage{upgreek}
\usepackage{todonotes}
\usepackage[colorlinks=true, allcolors=blue]{hyperref}
\usepackage{soul}

\begin{document}
\title{Directly visualizing the energy level structure of quantum dot molecules}

\author{Heun Mo Yoo}
\affiliation{Department of Physics and Astronomy, University of California, Los Angeles, California 90095, USA}
\author{Tanner M. Janda}
\affiliation{Department of Physics and Astronomy, University of California, Los Angeles, California 90095, USA}
\affiliation{Center for Quantum Science and Engineering, University of California, Los Angeles, California 90095, USA}
\author{Connor Nasseraddin}
\affiliation{Department of Physics and Astronomy, University of California, Los Angeles, California 90095, USA}
\affiliation{Center for Quantum Science and Engineering, University of California, Los Angeles, California 90095, USA}
\affiliation{Department of Electrical and Computer Engineering, University of California, Los Angeles, California 90095, USA}
\author{Jason R. Petta}
\email{petta@physics.ucla.edu}
\affiliation{Department of Physics and Astronomy, University of California, Los Angeles, California 90095, USA}
\affiliation{Center for Quantum Science and Engineering, University of California, Los Angeles, California 90095, USA}
\affiliation{Department of Electrical and Computer Engineering, University of California, Los Angeles, California 90095, USA}
\affiliation{HRL Laboratories, LLC, 3011 Malibu Canyon Road, Malibu, California 90265, USA}

\begin{abstract}
The orbital, spin and valley degrees of freedom in silicon quantum dots support many modes of spin qubit operation. However, it is generally challenging to obtain information about the energy level spectrum over large ranges of parameter space. We demonstrate a form of spectroscopy that is capable of mapping the energy level structure of a double quantum dot as a function of level detuning, interdot tunnel coupling, and magnetic field. In the one-electron regime, we directly observe the transition from the atom-like energy levels of isolated quantum dots to molecular-like bonding and anti-bonding states with increasing interdot tunnel coupling. We also resolve the Zeeman splitting of ground and excited valley states in a magnetic field. In the two-electron regime, we gain access to the detuning dependent singlet–triplet splitting. Our work may be extended to a broader class of systems, such as strong spin-orbit materials or proximitized quantum dots, allowing the direct extraction of various energy gaps. 
\end{abstract}

\maketitle
\renewcommand{\figurename}{\textbf{Fig.}}
\renewcommand{\thefigure}{\textbf{\arabic{figure}}}

Semiconductor spin qubits have many modes of operation, as the quantum dot (QD) energy levels depend strongly on the charge configuration \cite{Loss1998,Petta2005,Medford2013}, interdot tunnel coupling \cite{Oosterkamp1998,Petta2004,Koski2020}, and connectivity of the QDs within the device \cite{Burkard2023RMP}. A single spin qubit is realized in a single QD, where nanoscale confinement produces a discrete energy spectrum, and a static magnetic field lifts the spin degeneracy \cite{Loss1998,Koppens2006}. When two QDs are strongly tunnel coupled, molecular bonding and anti-bonding levels emerge, enabling the charge \cite{Hayashi2003,Petta2004} and flopping-mode qubits \cite{Benito2019b,Croot_flopping-mode_2020}. In coupled QDs, the exchange interaction between electrons further modifies the energy level structure of few-electron spin states, allowing the encoding of additional types of spin qubits, such as singlet–triplet \cite{Petta2005,Shulman2012,Nichol2017} and hybrid qubits \cite{Kim2014,Shi2014,Cao2016}.

Early experiments probed the excited states of single \cite{Kouwenhoven1997,Sapmaz2005} and double \cite{ono_current_2002,Sapmaz2006,Thelander2018} QDs using transport spectroscopy. However, this method requires strong tunnel coupling of the QDs to the reservoirs, a regime far from typical qubit operating conditions that demand a high degree of isolation from the environment to prevent decoherence. Pulsed-gate spectroscopy \cite{elzermanspec2004}, which relies on the charge detection of tunneling electrons, probes the energy levels of a QD nearly isolated from the reservoirs. However, its application so far remains limited to a single QD. More recent approaches, such as microwave spectroscopy \cite{Mi_PRL_2017}
and detuning axis pulsed spectroscopy \cite{chen_detuning_2021}, only provide information in narrow regions of parameter space near energy level crossings. As high fidelity quantum control of spin qubits requires detailed knowledge of the energy level structure, there is demand for a straightforward spectroscopy method that provides information over a wide range of parameter space.

In this manuscript, we demonstrate molecular energy level spectroscopy of a tunnel coupled double quantum dot (DQD) \cite{eriksson_paper}. Our approach allows the direct visualization of the energy level spectra of one- and two-electron regimes as a function of level detuning. The measured spectra provide a comprehensive view of the DQD energy levels, including the transition from atomic- to molecular-like states, the magnetic field dependence of spin and valley states, and the effects of the Pauli exclusion principle. Since the approach only relies on tunneling and sensitive charge detection, we anticipate it may be applied to other systems, such as hybrid quantum devices designed to host Majorana zero modes \cite{MZM_review}.

\begin{figure*}[tbh!]
	\centering
	\includegraphics[width=2\columnwidth]{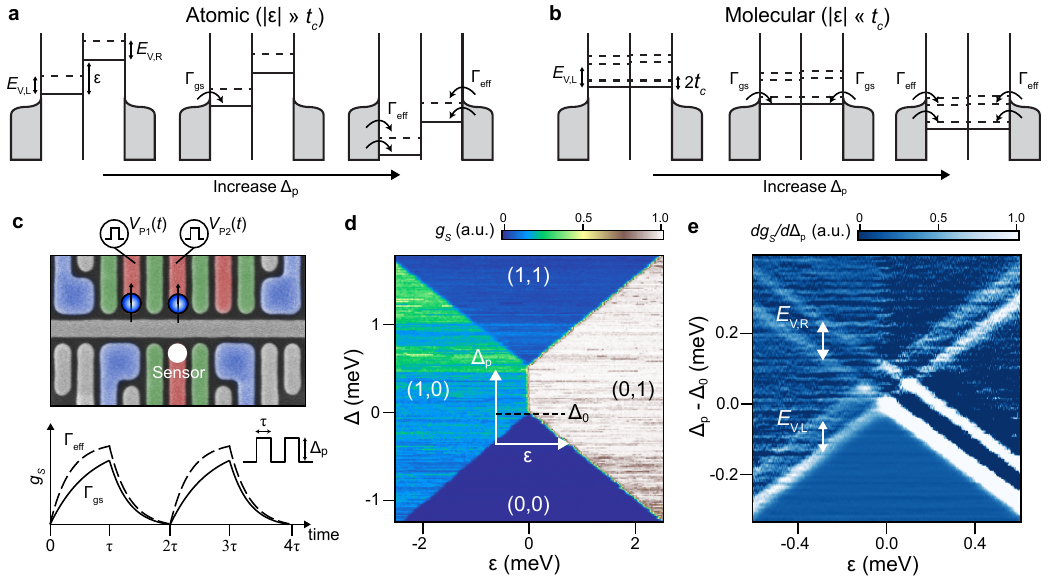}
	\caption{\textbf{Molecular spectroscopy.} \textbf{a},\textbf{b}, Level diagrams illustrating the spectroscopy of a DQD in both the atomic (\textbf{a}) and molecular (\textbf{b}) regimes. Both QDs are initially empty. Lowering their chemical potentials allows an electron to tunnel into the ground state, localized in one of the QDs at $|\varepsilon| \gg t_{c}$ \cite{Oosterkamp1998}. At small detuning $|\varepsilon| \ll t_{c}$, the degenerate states hybridize into bonding and anti-bonding states separated by $2t_{c}$ \cite{Oosterkamp1998,Petta2004}. Further lowering of the chemical potentials permits tunneling into either the ground or the excited states. \textbf{c}, False-color scanning electron microscope image of the device. Square waves are applied to the plunger gates, P1 and P2. The tunneling of electrons on and off the DQD is detected by measuring the charge sensor conductance $g_{s}$. Bottom panel: Time-averaged $g_{s}$, illustrating the characteristic $RC$ response. When the excited state becomes accessible, the loading rate increases. The inset shows the applied square wave with amplitude $\Delta_{p}$ and period $2\tau$. \textbf{d}, DQD stability diagram measured as a function of $\varepsilon$ and $\Delta$ with no pulses applied. White axes indicate the range of $\varepsilon$ and $\Delta_{p}$ used for the spectroscopy measurement in \textbf{e}. $\Delta_{0}$ represents the reference level at which electron loading into the lower triple point occurs. \textbf{e}, Single-electron spectrum showing the detuning dependence of the energy levels. The vertical axis $\Delta_{p}$ is offset by $\Delta_{0}$. At $\epsilon$ = 0, the ground states of the left and right QDs are degenerate.
    }
	\label{fig:1}
\end{figure*}

We perform molecular spectroscopy of a DQD formed beneath plunger gates P1 and P2 in an Intel Si/SiGe triple quantum dot device (Fig.~\ref{fig:1}c). In the absence of a magnetic field, the DQD can be modeled as a four-level system consisting of the lowest (ground) orbital of the left and right QDs, each split by the valley splitting (denoted as $E_{\rm V,L}$ and $E_{\rm V,R}$) \cite{Burkard2016,Zhao2022}. We tune the total electrochemical potential of the DQD through a virtualized gate defined as $\Delta = -(\mu_{L}+\mu_{R})/2$, where $\mu_{L(R)}$ is the electrochemical potential of the left (right) QD. The interdot detuning, defined as $\varepsilon = \mu_{L} - \mu_{R}$, is likewise controlled by a virtual gate orthogonal to the $\Delta$ axis (see Methods for details).

The experiment begins by setting $\Delta$ deep into the (0,0) charge state. We then apply a repeating square pulse, $\Delta(t)$, to modulate the electrochemical potential of both QDs. During the high phase of the pulse sequence, an electron tunnels from the reservoirs into the ground state of the DQD, which is localized in the atomic regime ($|\varepsilon|\gg t_{c}$) and delocalized in the molecular regime ($|\varepsilon|\ll t_{c}$). Here, $t_c$ denotes the interdot tunnel coupling (see Figs.~\ref{fig:1}a and b). During the low phase of the pulse sequence, the electron tunnels off of the DQD. To detect the tunneling response, we measure the conductance through a charge sensor, $g_{s}$ \cite{mills2022high,Zajac2016}. As an electron tunnels into and out of the DQD ground state in response to the square pulse, the time-averaged $g_{s}$ exhibits a characteristic $RC$-circuit response with a rate $\Gamma_{gs}$ determined by the tunnel coupling to the reservoirs (see the bottom panel of Fig.~\ref{fig:1}c) \cite{elzermanspec2004}. When the excited state becomes energetically accessible at larger pulse amplitudes, the effective tunneling rate $\Gamma_{\rm eff}$ increases, resulting in a faster $RC$ response and a change in the time-averaged $g_{s}$ \cite{Zajac2016}. We set the pulse duration $\tau$ = 0.3 -- 1 \textmu s, while the tunneling timescale (controlled via the reservoir–dot barrier gates) is set to $\Gamma_{gs}^{-1} \sim$ 50 \textmu s. 

We first perform spectroscopy in the one-electron regime at zero magnetic field ($B$ = 0) to examine the orbital and valley degrees of freedom. In Fig.~\ref{fig:1}d, the DQD charge stability diagram is plotted as a function of $\varepsilon$ and $\Delta$ in the absence of pulses. We set $\Delta = \text{-0.34 meV}$ and $\varepsilon=\text{0 meV}$ to deplete the DQD and then apply a voltage pulse with magnitude $\Delta_p$ to probe the one-electron excited states, where the height of $\Delta_p$ is proportional to the energy at which an electron can tunnel into the DQD (see Figs.~\ref{fig:1}a and b). In our measurements, we define the energy relative to the ground state energy, $\Delta_0$, measured at zero detuning. By taking a numerical derivative of the sensor signal and varying $\varepsilon$, we acquire a $dg_{s}/d \Delta_{p}$ map representing the one-electron energy level spectrum.


The energy level spectrum in Fig.~\ref{fig:1}e shows both the ground and excited states of the DQD at weak interdot tunnel coupling. The ground state traces out an inverted V shape, consistent with the (0,0) $\leftrightarrow$ (0,1) and (0,0) $\leftrightarrow$ (1,0) charge transition lines in the stability diagram. The upper-valley states reside at energies $E_{V,L}=89\pm5$ {\textmu}eV and $E_{V,R}=107\pm6$ {\textmu}eV above the ground state, for the left and right QDs, respectively. As $\varepsilon$ is varied, these valley states move diagonally and cross without coupling to each other.

\begin{figure}[tb!]
	\centering
	\includegraphics[width=\columnwidth]{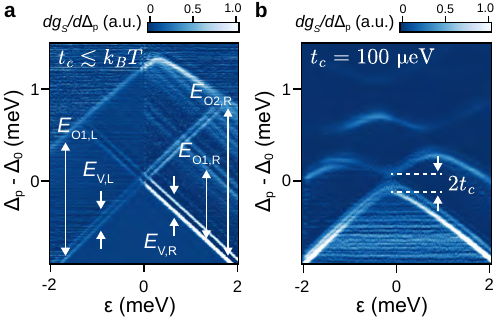}
	\caption{\textbf{Transition from artificial atom to molecule.} \textbf{a},\textbf{b}, One-electron spectra of the DQD at (\textbf{a}) weak and (\textbf{b}) strong $t_{c}$. $E_{O1,L}$ and $E_{O1,R}$ denote the first orbital excited state energies in the left and right QDs, respectively. $E_{O2,R}$ is the energy of the second orbital excited state in the right QD. At small $t_{c}$, the orbitals of the left and right QDs are localized. As $t_{c}$ increases, the localized orbitals hybridize to form bonding and anti-bonding states \cite{Oosterkamp1998}, as indicated by the avoided crossings with a splitting of $2t_c$ in panel \textbf{b}.}
	\label{fig:2}
\end{figure}

We next investigate the evolution of the one-electron energy levels as interdot tunnel coupling is increased and the device transitions to the molecular bonding and anti-bonding regime. Figure~\ref{fig:2} compares the one-electron spectra taken at weak coupling $(t_{c} \lesssim k_{B}T)$ and strong coupling $(t_{c}$ = 100 {\textmu}eV), measured over a broader range of $\varepsilon$ and $\Delta$. The weak coupling spectrum shows the first and second orbital excited states in the right QD at energies $E_{O1,R}=1.62\pm0.08 \textrm{ meV}$ and $E_{O2,R}=2.93\pm0.15 \textrm{ meV}$ above the ground state. At negative detuning, the left dot orbital excited state energy $E_{O1,L}=0.78\pm0.04 \textrm{ meV}$ is readily extracted. Similar to the level crossings in Fig.~\ref{fig:1}d, both the left and right dot energy levels pass through each other without hybridization.

At large tunnel coupling, a qualitatively different energy level structure emerges as the electron spreads out over the DQD, forming bonding and anti-bonding states \cite{Oosterkamp1998}. The spectrum in Fig.~\ref{fig:2}b exhibits a host of avoided crossings. The coupling strength deduced from the avoided crossing in the ground state orbital is $97\pm6$~{\textmu}eV, in close agreement with $t_{c}$ = 100 {\textmu}eV extracted from charge sensing measurements taken along the (0,1) $\leftrightarrow$ (1,0) interdot transition (see Extended data Fig. 2). As expected, the energy gaps between bonding and anti-bonding states in the higher orbitals are larger than that of the ground state orbital since the excited state wavefunctions penetrate deeper into the interdot barrier. 

\begin{figure}[tb!]
	\centering
	\includegraphics[width=\columnwidth]{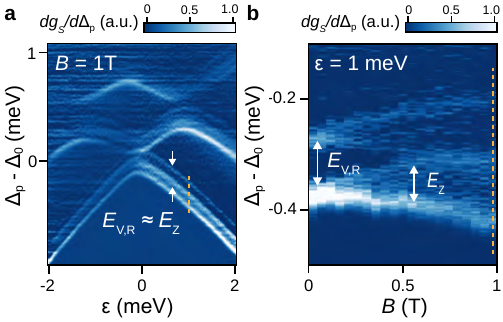}
	\caption{\textbf{Zeeman splitting.} \textbf{a}, One-electron spectrum taken at $B$~=~1~T, showing Zeeman splitting $E_{Z}$ of the lowest orbital. The orange dashed line indicates the value of detuning used to acquire the data in panel \textbf{b}. \textbf{b}, The level spectrum measured as a function of $B$ with $\varepsilon$~=~1~meV. The orange dashed line corresponds to $B$~=~1~T, the field at which the panel \textbf{a} data are acquired. From the level splitting of the lower- and upper-valley states, we extract $g$ = 1.98 $\pm0.12$.} 
	\label{fig:3}
\end{figure}

In the one-electron regime, we can directly measure the Zeeman splitting of the energy levels as a function of magnetic field $B$. With $B$ = 1 T, the spectrum shows an additional energy level arising from the Zeeman splitting of the valley states in the lowest orbital (see $E_{Z}$ indicated by the white arrows in Fig.~\ref{fig:3}a). The Zeeman splitting is less pronounced at negative detuning, as the left QD is located further away from the charge sensor reducing the signal-to-noise ratio. To quantitatively analyze the Zeeman splitting and extract the electronic $g$-factor, we measure the spectrum as a function of $B$ with $\varepsilon$ = 1 meV (see Fig.~\ref{fig:3}b). We clearly see the valley states in the right QD split into four energy levels as $B$ increases and lifts the spin degeneracy. At $B$~$\approx$~0.9~T, when $g\mu_{B}B$~$\approx$~$E_{V,R}$, the spin-up lower-valley state becomes degenerate with the spin-down upper-valley state. By fitting the Zeeman splitting of the lower- and upper-valley states, we find $g$~=~$1.98\pm0.12$, which agrees well with the free electron $g$-factor \cite{roth_g_1960} (see Extended data Fig. 3).

\begin{figure*}[htb!]
	\centering
    \includegraphics[width=2\columnwidth]{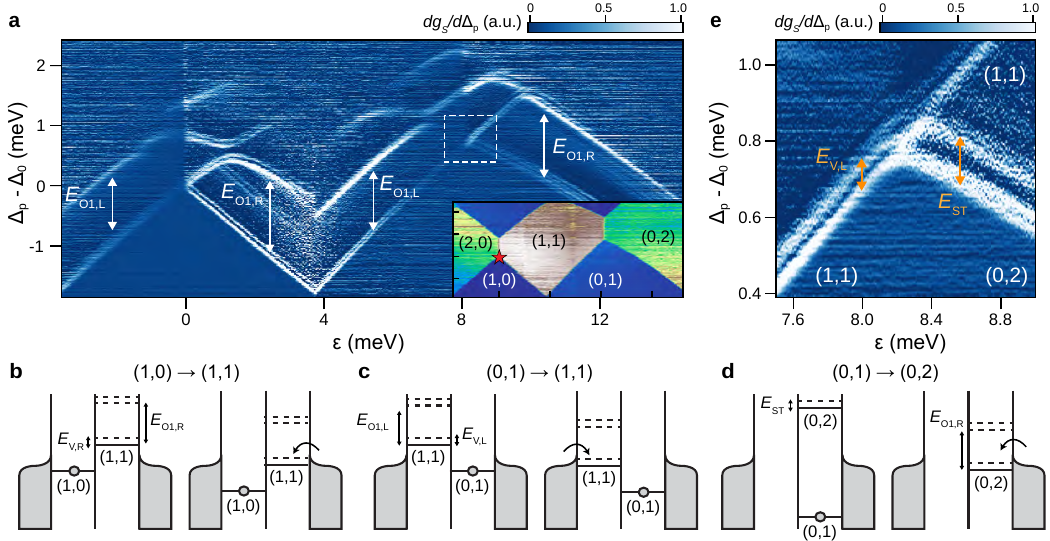}
	\caption{\textbf{Two-electron state spectroscopy}. \textbf{a}, Two-electron spectrum acquired at $B$~=~0~T, showing the ground and excited states of the (2,0), (1,1) and (0,2) charge configurations. The detuning axis $\varepsilon$ is measured relative to the lower (2,0)~$\leftrightarrow$~(1,1) triple point. The vertical axis $\Delta_p$ is offset by the reference level $\Delta_0$, corresponding to the energy at the same triple point (see the red star in the inset). The stability diagram in the inset shows the spectroscopy measurement window. \textbf{b}--\textbf{d}, Cartoons illustrating tunneling processes in the two-electron regime. At small positive $\epsilon$ (\textbf{b}), a single electron initially resides in the left QD. A $\Delta$ pulse injects an extra electron into the right QD. The spectrum, in this case, reveals the valley and orbital excited states of the right QD. At a larger $\epsilon$ (\textbf{c}), the DQD is in the (0,1) configuration with an electron occupying the right QD. The excited state spectrum is acquired by adding a second electron into the left QD. At even larger $\epsilon$ (\textbf{d}), the DQD remains in the (0,1) configuration, but the second electron can tunnel into the (0,2) singlet or triplet state. \textbf{e}, High resolution spectrum acquired near the (1,1)~$\leftrightarrow$~(0,2) transition. In the (1,1) charge configuration, the ground state consists of two electrons occupying the lower-valley states of the left and right QDs. In the excited state, one electron occupies the upper-valley state \cite{Philips2022}. As $\epsilon$ increases, these (1,1) states moves up in energy and cross the singlet and triplet states in the (0,2) charge configuration. $E_{ST}$ denotes the energy gap between the singlet and triplet states.}
	\label{fig:4}
\end{figure*}

The earliest spin qubit demonstrations were performed in the GaAs/AlGaAs material system \cite{Petta2005,Burkard2023RMP}. In contrast to GaAs, silicon possesses an additional valley degree of freedom that can compromise the operation of spin qubits \cite{Yang2013}. We therefore investigate the two electron regime by accumulating one electron in the DQD and then applying a $\Delta$ pulse to acquire the addition spectrum for a second electron loaded into the DQD. The inset of Fig. \ref{fig:4}a shows the DQD charge stability diagram in the vicinity of the two-electron charge states (2,0), (1,1), and (0,2). Prior to applying the $\Delta$ pulse, we initialize the DQD in either the (1,0) or (0,1) charge state.

Figure~\ref{fig:4}a shows the excited state spectrum of the two-electron DQD at $B=\text{0 T}$. The detuning axis is referenced to the (2,0) ${\leftrightarrow}$ (1,1) interdot charge transition (red star in the inset of Fig.~\ref{fig:4}). Similar to the one-electron spectrum, the two-electron spectrum shows the orbital energies of the left and right QDs, valley splittings, and avoided crossings at high energies. Starting from negative $\varepsilon$, we identify the ground and first orbital excited states of the left QD separated by $E_{O1,L}$. Between $\varepsilon=\text{0 meV}$ and 4 meV, we observe the valley splitting $E_{V,R}$ and the higher orbitals of the right QD. Near $\varepsilon = \text{4 meV}$, we find a discontinuous shift in the energy levels. This discontinuous shift occurs because we probe the left QD's excited states rather than those of the right QD at $\varepsilon$~$\gtrsim$~4~meV, as the spectrum is acquired by adding a second electron into the left QD (see Figs.~\ref{fig:4}b and c for a detailed explanation). At $\varepsilon> \text{8.2 meV}$, we identify the right QD's ground and first excited orbital states since the spectrum is measured by adding an extra electron to the right QD (see Fig.~\ref{fig:4}d).

To study the two-electron singlet and triplet states, we perform a high-resolution scan near the (1,1) $\leftrightarrow$ (0,2) interdot transition (see Fig.~\ref{fig:4}e). In the (1,1) charge state, the ground state corresponds to two electrons occupying the lower-valley states in the left and right QDs, while the excited state consists of one electron occupying the upper-valley state and another electron occupying the lower-valley state. As $\varepsilon$ increases, these ground and excited states transition into singlet and triplet states in the (0,2) charge configuration, with the (1,1) states moving up in energy with detuning. Since the (0,2) triplet state arises from one electron occupying the upper-valley state of the right QD, the singlet-triplet energy gap $E_{ST}$ is given by $E_{V,R}$ (see Fig.~\ref{fig:4}d). From the spectrum, we find $E_{ST} =98\pm6$ \textmu eV, which agrees well with the $E_{V,R}$ measured in Fig.~\ref{fig:1}e (see also Extended data Table 1 for the valley splittings measured in the (1,1) configuration). 

Finally, we probe the singlet-triplet splitting arising from the exchange interaction in the left QD. The spectrum for the (2,0) and (1,1) charge configurations in Extended data Fig.~4 shows a qualitatively similar trend to that observed in Fig.~\ref{fig:4}e. For instance, the valley-split states in the (1,1) configuration cross the (2,0) singlet and triplet states near the interdot transition. Furthermore, we observe a small $E_{ST}=69\pm4$~{\textmu}eV, comparable to $E_{V,L}=64\pm4$~{\textmu}eV measured in the (1,1) charge state, which is consistent with the spin blockade picture of $E_{ST}$ limited by valley splitting in a two-electron QD \cite{Philips2022}.

In summary, we have demonstrated a new approach for directly visualizing the energy level structure of quantum dot molecules. Our measurements reveal clear signatures of orbital, valley, and Zeeman physics. Avoided crossings and singlet-triplet splittings are observed, providing useful parameters in the Hamiltonians that govern spin qubit operation. Moreover, the present work can be expanded to investigate multi-electron QDs, where electron-electron interactions produce energy levels that deviate from a noninteracting model \cite{Philips2022} and impact the control and readout of spin qubits \cite{Abadillo-Uriel2p21}. We envision the broad applicability of this simple measurement approach to characterize the energy level structure of QDs as well as to provide insights into less mature materials systems designed to support Majorana zero modes \cite{MZM_review}.

Note added: While preparing this manuscript, we became aware of concurrent work \cite{eriksson_paper} reporting the energy spectrum of a double quantum dot in the (1,3) and (0,4) configuration.

\begin{acknowledgments}
Research was sponsored by the Army Research Office and was accomplished under Cooperative Agreement No.~W911NF-22-2-0037 and Grant No.~W911NF-23-1-0104. The views and conclusions contained in this document are those of the authors and should not be interpreted as representing the official policies, either expressed or implied, of the Army Research Office or the U.S. Government. The U.S. Government is authorized to reproduce and distribute reprints for Government purposes notwithstanding any copyright notation herein. We acknowledge support from Intel Corporation for providing the device, as well as technical conversations with Nathan Bishop, Joelle Corrigan, Matthew Curry, and Rene Otten.
\end{acknowledgments}

\section*{Methods}
\subsection*{Device}
The triple quantum dot (TQD) was fabricated using an isotopically purified $^{28}$Si quantum well (800 ppm residual $^{29}$Si). Details of the gate structure used to define the TQD and charge sensor are described in Ref.~\cite{Neyens2024}. A DQD is formed under two plunger gates P1 (left QD) and P2 (right QD), while the reservoirs are formed under the third plunger gate and the accumulation gates (shaded in blue in Fig.~\ref{fig:1}c). The tunnel coupling $t_c$ is tuned via the interdot barrier gate (see the green gate between P1 and P2 in Fig.~\ref{fig:1}c). The sensor dot is separated from the TQD array by the center screening gate.  Since the sensor is directly across from P2, the charge sensitivity is larger for the right QD \cite{Zajac2016}.

\subsection*{Measurement and calibration procedure}
To independently control $\Delta$ and $\epsilon$, we define virtual gates that compensate for the cross-capacitance in the device \cite{Mills2019}. We define $\Delta =\alpha({u_{\rm P1}+u_{\rm P2}})/2$ and $\epsilon$~=~$\alpha({u_{\rm P2}-u_{\rm P1}})$, where $u_{\rm P1}$ and $u_{\rm P2}$ are virtualized plunger gates proportional to $-\mu_{L}/e$ and $-\mu_{R}/e$. The lever arm $\alpha$ is calibrated via bias triangle measurements. Furthermore, we tune the reservoir-to-dot barrier gates to set the tunneling rate $\Gamma$ $\sim0.02$ \textmu $\text{s}^{-1}$, which is estimated from real-time electron hopping measurements taken at the charge transitions of both left and right QDs \cite{Zajac2016}. The reservoir-to-dot barrier gates are held fixed for each spectroscopic measurement, except for the two-electron excited-state measurement, where $\epsilon$ is varied over a wide range. In Fig.~\ref{fig:4}a, we adjust the virtualized reservoir-to-dot barrier gates near the (1,0) $\leftrightarrow$ (0,1) interdot transition to maintain a nearly constant tunneling rate. The stability diagram shown in the inset of Fig.~\ref{fig:4}a is taken while tuning the virtualized reservoir-to-dot barrier gates near the interdot transition.

We measure the sensor dot conductance $g_{s}$ using a 50~\textmu V excitation at 400 kHz (see Ref.~\cite{mills2022high} for details of the readout circuit). The excited state spectra are then acquired by applying $\Delta$ pulses and recording $g_{s}$ and its numerical derivative $dg_{s}/d\Delta_{p}$. We note that a Gaussian low-pass filter is applied to the measured spectra to smooth out the noise in $dg_{s}/d\Delta_{p}$. In order to ensure that no distortion is introduced by filtering, we use a Gaussian filter with a standard deviation $\sigma$ smaller than the thermal broadening. Extended data Fig.~1a shows the raw $g_{s}$ data, measured as a function $\varepsilon$ and $\Delta_{p}-\Delta_{0}$, for the one-electron spectrum presented in Fig.~\ref{fig:1}e. There are no discernible differences in the ground and excited states between the raw $g_{s}$ data and the spectrum in Fig.~\ref{fig:1}e.

\subsection*{Spectral resolution}
When the reservoir-to-dot tunneling rate $\Gamma$ is sufficiently small, the energy resolution is limited by thermal broadening, as an electron is injected from thermally broadened reservoirs. Extended data Fig.~1b shows a Fermi function \cite{Zajac2016} fitted to the raw $g_{s}$ data. From the Fermi function fit, we extract a temperature of $T=~94~\text{mK}$. We also confirm that the full width at half maximum (FWHM) of the $dg_{s}/d\Delta_{p}$ peak approximately corresponds to the thermal broadening $\sim 3.5 k_{B}T$ (see Extended data Fig.~1c).

\subsection*{Charge sensing measurement of tunnel coupling}
To compare with the energy level spectra, we can also extract the interdot tunnel coupling $t_{c}$ using the charge sensing measurement introduced by DiCarlo \textit{et al.} \cite{dicarlospec2004}. Extended data Fig.~2 shows $g_{s}$ measured as a function of $\varepsilon$ near the (0,1) $\leftrightarrow$ (1,0) interdot charge transition (no pulses applied). The values of $t_{c}$ extracted from the two-level system model \cite{dicarlospec2004,Zhao2022} are 11~\textmu eV for weak coupling and 100~\textmu eV for strong coupling. For weak coupling, the value of $t_{c}$ used in the fit is comparable to the thermal broadening of a Fermi function. Thus, we label weak coupling as $t_{c} \lesssim k_{B}T$.

\subsection*{Electronic $g$-factor}
To extract the $g$-factor, we fit the $B$-dependent data of Fig.~\ref{fig:3}b with Gaussian functions. Extended data Fig.~3a shows vertical cuts of the spectrum in Fig.~\ref{fig:3}b. We fit the two peaks corresponding to the spin-down lower-valley and spin-up upper-valley states using Gaussian functions (see Extended data Fig. 3a). We then extract the $g$-factor from their energy difference, corresponding to $E_{Z}+E_{V,R}$, measured as a function of $B$ (see Extended data Fig.~3c). The extracted value of the $g$-factor is 1.98. We estimate the uncertainty in the measured $g$-factor to be $\pm~0.12$ based on a $\sim 6 \%$ scaling error in the pulse amplitude calibration (see the next section for details).

\subsection*{Uncertainty due to $\Delta_{p}$ calibration error}
We attribute the main source of error in our measurements to the pulse amplitude calibration. We calibrate pulse amplitudes $V_{P1}$ and $V_{P2}$ based on gate voltage offsets observed in the charge transitions with the pulses applied to the DQD \cite{Petta2005b}. After calibration, we find a few percentage error in $\Delta_{p}$ that varies with gate voltages and pulse width $\tau$. We estimate the $\Delta_{p}$ error by comparing the two sets of data: stability diagrams measured with dc voltage sweeps and ground state energy levels measured with pulse amplitude sweeps.

Extended data Fig.~5b shows the (0,0) $\leftrightarrow$ (0,1) and (0,0) $\leftrightarrow$ (1,0) charge transition lines measured by sweeping dc voltage $\Delta$ (blue) and pulse amplitude $\Delta_{p}$ (red) at weak $t_c$. We observe a misalignment between the two datasets. Since rescaling the pulse amplitude by 5$\%$ corrects this misalignment, we estimate the error to be $\pm~\Delta_{p}\times 5\%$ in the weak $t_{c}$ spectrum. The errors in $\Delta_{p}$ for other spectra are estimated using the same method.

\newpage
\setcounter{figure}{0}
\renewcommand{\figurename}{\textbf{Extended Data}}
\renewcommand{\thefigure}{\textbf{Fig. \arabic{figure}}}

\begin{figure}[tb!]
	\centering
    \includegraphics[width=\columnwidth]{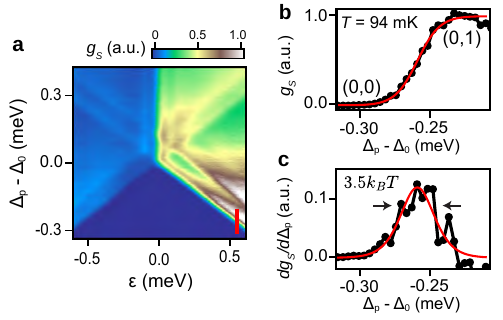}
	\caption{\textbf{Raw charge sensor conductance and spectral broadening.} \textbf{a}, Raw $g_{s}$ data for the one-electron spectrum presented in Fig.~1e. The raw data clearly exhibit the valley states of the left and right QDs. The charge sensing contrast is smaller for the left QD, as it is located further away from the charge sensor. \textbf{b}, Broadening of the (0,0) $\leftrightarrow$ (0,1) charge transition. The black circles represent a vertical cut of the $g_{s}$ data at $\varepsilon = \text{0.55 meV}$ (see the red line in \textbf{a}). The data are fit to a Fermi function with $T = \text{94 mK}$ (red line). \textbf{c}, The derivative $dg_s/d\Delta_{p}$ data (black circles) fitted to the derivative of a Fermi function (red line). The full width at half maximum corresponds to $\sim 3.5 k_{B}T$.}.
	\label{fig:S1}
\end{figure}

\begin{figure}[tb!]
	\centering
    \includegraphics[width=\columnwidth]{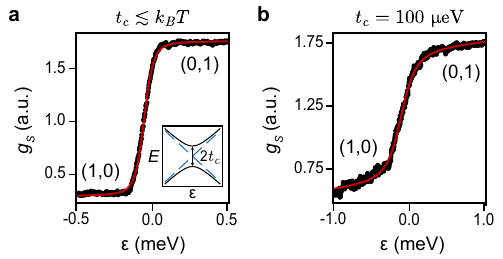}
	\caption{\textbf{Charge sensing measurement of the interdot tunnel coupling.} \textbf{a},\textbf{b}, $g_{s}$ measured across the (1,0) $\leftrightarrow$ (0,1) interdot transition at weak and strong $t_{c}$. The data (black circles) are fitted to the two-level system model \cite{dicarlospec2004} (red lines). At weak coupling in \textbf{a}, the broadening of the interdot transition is dominated by $k_{B}T$ (see Methods). On the other hand, the interdot transition in \textbf{b} is fitted with $t_c$ = 100 \textmu eV, which is greater than thermal broadening. The inset illustrates a `tunnel-coupled two-level system' \cite{dicarlospec2004}, where the ground and excited states are split by $2t_{c}$ at $\varepsilon$ = 0.}
	\label{fig:S2}
\end{figure}

\begin{figure}[tb!]
	\centering
    \includegraphics[width=\columnwidth]{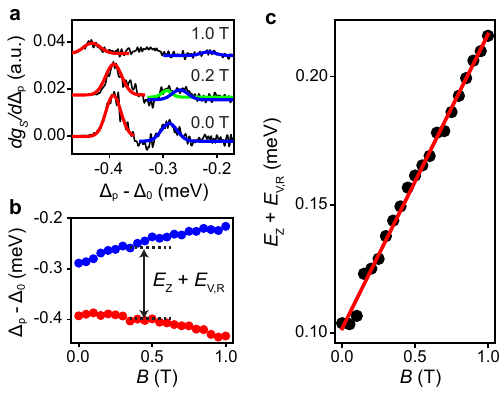}
	\caption{\textbf{$g$-factor measurement.} \textbf{a}, Tunneling spectra at $B=$ 0, 0.2, and 1 T while $\varepsilon$ is fixed at 1 meV. The black lines are vertical cuts from the data in Fig.~3b. Single Gaussian functions are fitted to the peaks associated with the spin-down lower-valley (red) and spin-up upper-valley states (blue). For overlapping peaks, the sum of two Gaussian functions is fitted (see the green and blue Gaussian curves at $B$ = 0.2 T). \textbf{b}, The $B$ dependence of the spin-up upper-valley (blue circles) and spin-down lower-valley (red circles) energies. The energy difference corresponds to $E_{Z}+E_{V,R}$. \textbf{c}, $g$-factor extracted from the $B$ dependence of $E_{Z}+E_{V,R}$ (black circles). The red line shows a linear fit $g\mu_{B}B+E_{V,R}$, where $\mu_{B}$ is the Bohr magneton. The fit parameters are $g=1.98\pm0.12$ and $E_{V,R}=102\pm6$~{\textmu}eV.}
	\label{fig:S3}
\end{figure}

\renewcommand{\tablename}{\textbf{Extended Data}}
\renewcommand{\thetable}{\textbf{Table \arabic{table}}}

\begin{table}[htb!]
     \caption{\label{tab:1} \textbf{Summary of energy gaps extracted from the excited-state spectra}.}
     \begin{ruledtabular}
     \begin{tabular}{ccc}
     & \begin{tabular}[c]{@{}c@{}}
     Charge state \end{tabular} & Energy (\textmu eV) \\
    \hline  $E_{V,L}$ & (0,1) and (1,0)&    $89\pm5$ \\
    $E_{V,L}$ & (1,1)&   $64\pm4$ \\
    $E_{V,R}$ & (0,1) and (1,0) & $107\pm6$ \\
    $E_{V,R}$ & (1,1)&   $101\pm5$ \\
    $E_{ST}$ & (2,0) &   $69\pm4$     \\
    $E_{ST}$ & (0,2) & $98\pm6$         \\
    $E_{O1,L}$ & (1,0) & $780\pm40$ \\
    $E_{O1,L}$ & (1,1) &  $1,180\pm60$\\
    $E_{O1,R}$ & (0,1) & $1,620\pm80$ \\
    $E_{O1,R}$ & (1,1) &  $1,370\pm70$\\
    $E_{O2,R}$ & (0,1) & $2,930\pm150$ \\   
\end{tabular}

     \end{ruledtabular}
 \end{table}
 
\begin{figure}[tb!]
	\centering
    \includegraphics[width=\columnwidth]{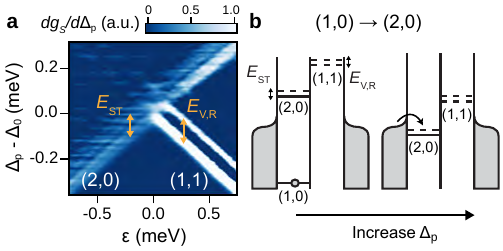}
	\caption{\textbf{Singlet-triplet energy splitting in the left quantum dot.} \textbf{a}, A high-resolution scan of the spectrum near the (2,0) $\leftrightarrow$ (1,1) interdot transition. At positive $\varepsilon$, the ground state is in the (1,1) charge configuration, consisting of two electrons occupying the lower-valley states of the left and right QDs. In the first excited state, one electron occupies the lower-valley state in the left QD while the other electron occupies the upper-valley state in the right QD (see Fig.~4b for a detailed explanation). At negative $\varepsilon$, the (2,0) configuration hosts the singlet ground state and triplet excited state, separated by $E_{ST}$. \textbf{b}, A cartoon describing the spectroscopy of the (2,0) excited state via electron injection into the DQD prepared in the (1,0) charge configuration. At negative $\varepsilon$, one electron initially resides in the left QD. A $\Delta$ pulse can load a second electron into the singlet state or higher-energy triplet state in the left QD.}
	\label{fig:S4}
\end{figure}

\begin{figure}[htb!]
	\centering
    \includegraphics[width=\columnwidth]{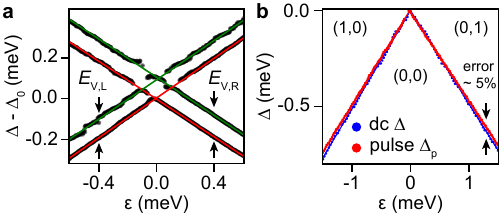}
	\caption{\textbf{Extraction of valley splittings and uncertainty in $\Delta_{p}$}. \textbf{a}, The valley-split energy levels in the (0,1) and (1,0) charge configuration. The black circles represent the $dg/d\Delta_{p}$ peaks in the spectra shown in Fig.~1e. The red and green lines are linear fits that represent the lower-valley and upper-valley states, respectively. The valley splittings are determined from the vertical offsets between the green and red lines. \textbf{b}, The (0,0) $\leftrightarrow$ (1,0) and (0,0) $\leftrightarrow$ (0,1) charge transition lines measured by sweeping $\Delta$ (blue) and $\Delta_{p}$ (red). A small pulse amplitude calibration error results in misalignment between the blue and red lines. The uncertainty in $\Delta_{p}$ due to the calibration error is estimated to be $\pm~5\%$.}
  	\label{fig:S5}
\end{figure}

\clearpage
\bibliography{bib_spectroscopy_v2,RMP_master2_bib_v4}

\end{document}